\documentclass[useAMS,usenatbib]{mn2e}

\begin{document}

%
\newcommand{\Hii}{\hbox{H\,{\sc ii}}}
\newcommand{\um}{\hbox{$\mu$m}}
\def\le{\mathrel{\hbox{\rlap{\hbox{\lower4pt\hbox{$\sim$}}}\hbox{$<$}}}}
\def\ge{\mathrel{\hbox{\rlap{\hbox{\lower4pt\hbox{$\sim$}}}\hbox{$>$}}}}
\def\gs{\mathrel{\raise1.16pt\hbox{$>$}\kern-7.0pt
\lower3.06pt\hbox{{$\scriptstyle \sim$}}}}
\def\ls{\mathrel{\raise1.16pt\hbox{$<$}\kern-7.0pt
\lower3.06pt\hbox{{$\scriptstyle \sim$}}}}
\title{The Near-Infrared Extinction Law in Regions of High $A_V$}
\author[T.\ J.\ T.\ Moore et al.] {T.\ J.\ T.\ Moore$^1$
        S.\ L.\ Lumsden$^{2}$
        N.\ A.\ Ridge$^{3}$
        P.\ J.\ Puxley$^4$\\
$^1$Astrophysics Research Institute, Liverpool John Moores University,
Twelve Quays House, Egerton Wharf, Birkenhead, CH41 1LD, UK.\\
$^2$Department of Physics and Astronomy, University of Leeds, Leeds LS2 9JT\\
$^4$Harvard-Smithsonian Center for Astrophysics, 60 Garden St, Cambridge, MA 02138 
USA\\
$^5$Gemini Observatory, c/o AURA, Casilla 603, La Serena, Chile }
%

\date{This is a preprint of an Article accepted for publication in Monthly Notices of the Royal Astronomical Society \copyright 2005 RAS}

\pagerange{\pageref{firstpage}--\pageref{lastpage}} \pubyear{2005}

\maketitle
\begin{abstract}
We present a spectroscopic study of the shape of the dust-extinction law
between 1.0 and 2.2\,$\mu$m towards a set of nine ultracompact
\hbox{H\,{\sc ii}} regions with $A_V \ge 15$\, mag. We find some evidence
that the reddening curve may tend to flatten at higher extinctions, but 
just over half of the sample has extinction consistent with or close to
the average for the interstellar medium.
There is no evidence of extinction curves significantly steeper than
the standard law, even where water ice is present.
Comparing the results to the predictions of a simple extinction model, we
suggest that a standard extinction law implies a robust upper limit to
the grain-size distribution at around $0.1 - 0.3$\,$\mu$m.  Flatter
curves are most likely due to changes in this upper limit,
although the effects of flattening due to unresolved clumpy
extinction cannot be ruled out.

\end{abstract}

\begin{keywords}
dust, extinction; H{\sc ii} regions; infrared: ISM
\end {keywords}
\section{Introduction}
\label{intro}
Accurate
knowledge of the extinction law is crucial to most forms of observational
astronomy.  A detailed understanding of extinction in the infrared (IR)
waveband, as it applies in differing environments, is especially important for
the study of any heavily embedded object, from young stars
to the cores of starburst galaxies and AGN.  The
average extinction law in the diffuse interstellar medium is well defined in
the near-IR and apparently shows little variation (e.g.\ Mathis 1990).
However, the broad studies which produce this conclusion are based on
low-extinction data (e.g.\ Savage \& Mathis 1979 with $A_V \ls 2$; Cardelli, 
Clayton \& Mathis 1989 with $A_V \ls 5$). There
is a little accurate information on how extinction curves might vary along
lines of sight to heavily embedded objects.  It is this latter question that we
are investigating in this paper.

There are many reasons to expect an altered extinction curve
in regions of high gas density and at large A$_V$.
Average grain sizes are expected to be higher in sheltered molecular cloud
environments than in the diffuse interstellar medium because of the processes
of grain coagulation and the growth of icy mantles where A$_V \ge 3$
(Whittet et al.\ 1983, Tielens 1989).  There
is now plenty of evidence for significantly altered grain-size distributions
and increased average grain sizes in dense molecular clouds and star-formation
regions. Such alterations cause variations in the extinction
parameter $R = A_V/E(B-V)$ and the wavelength of
maximum polarisation by dichroic absorption
where the extinction is higher (Savage \& Mathis 1979, Whittet et al.\ 2001).
In general, the value of $R$ rises from the interstellar
average of 3.1 to between 4 and 5 in dark clouds and regions of
star formation.  Such a change indicates a flattening of the
extinction curve at wavelengths shorter than 0.5\,$\mu$m and
is associated with alterations in the population of smaller
dust grains (e.g. Cardelli \& Clayton 1991) .

Additional evidence for grain processing in dense regions comes from, e.g.,
IR reflection
nebulae associated with star formation, in the wavelength dependence of the
scattered intensity (Castelaz et al.\ 1985), the scattering optical depth 
(Yamashita et al.\ 1989)
and the shape of the 3.09\,$\mu$m ice absorption feature (Pendleton, Tielens
\& Werner 1990). All appear to be affected by the presence of
significant numbers of large dust grains, and depletion of smaller grains may
cause changes in the near-IR continuum extinction in low-mass YSOs
(Moore \& Emerson 1994).

We have carried out observations of hydrogen recombination lines
in a number of compact and ultra\-compact \hbox{H\,{\sc ii}} regions,
and calculated the extinction
law for these objects by comparing the line ratios with the
predictions of well-established theoretical models (Hummer \& Storey 1995).
This method has been used previously by Landini et al.\ (1984) and
Lumsden \& Puxley (1996).  The latter found that the
G45.12+0.13 \hbox{H\,{\sc ii}} region displays an extinction law
somewhat steeper than the standard curve.  This paper extends the latter study,
looking for evidence of widespread variations from standard extinction
towards ultracompact H{\sc ii} regions with $A_V \ge 15$.

\section{Observations}
\label{obs}

\begin{table*}
\begin{minipage}{130mm}
\begin{centering}
\caption{Targets, observed positions and slit orientations}
\label{obstable}
\begin{tabular}{|l|l|l|l|c|l|}
Object &  IRAS name & RA (J2000)  & Dec. (J2000) & Distance$^{a}$ (kpc) & slit angle$^b$ \\
\hline
G29.96--0.02 & 18434--0242 & 18 46 03.9 & $-02$ 39 21.9 & 6.0 & 60$^\circ$ \\
G35.20--1.74 & 18592+0108 & 19 01 46.5 & +01 13 24.8 & 3.0 & 90$^\circ$ \\
G43.89--0.78 & 19120+0917 & 19 14 26.2 & +09 22 33.9 & 9.0 & 90$^\circ$  \\
G45.45+0.06 & 19120+1103 & 19 14 21.4 & +11 09 14.1 & 4.2/8.8 & 90$^\circ$  \\
W51d & 19213+1424 & 19 23 39.9 & +14 31 08.3 & 7.0 & 121$^\circ$  \\
RCW\,57 & 11097--6102 & 11 11 54.0 & $-61$ 18 23.9 & 3.1 & 90$^\circ$  \\
G298.23--0.33 & 12073--6233 & 12 10 00.9 & $-62$ 49 54.5 & 11.0 & 90$^\circ$  \\
G307.57--0.62 & 13291--6249 & 13 32 31.2 & $-63$ 05 19.8 & 2.7 & 90$^\circ$  \\
G337.95--0.48 & 16374--4701 & 16 41 08.1 &  $-47$ 06 46.7 & 2.9 & 90$^\circ$  \\
He2-117 & & 13 42 36.3 & $-61$ 22 26.6 & -- & 90$^\circ$  \\
He2-96 & & 15 05 59.5 & $-55$ 59 18.5 & -- & 90$^\circ$  \\
\hline
\end{tabular}
\end{centering}
$^a$See \S3.2 for references;
$^b$East of north
\end{minipage}
\end{table*}

Observations were made on 18, 19 and 20 May 1997 and 10 and 12 June 1998 with
the common-user infrared imaging spectrometer IRIS on the AAT with the cs/36
secondary.  Nine H{\sc ii} regions were observed with two planetary nebulae as
comparison sources.  Standards were observed at similar airmass, and where
possible G-type stars were used, as they have few intrinsic absorption
features. Where F-type stars had to be used, we interpolated over absorption
lines in their spectra.

Observations were made in two modes with IRIS.  The main data presented here
were obtained using cross-dispersed echelle grisms to obtain complete coverage
of {\em IJ} or {\em HK} in a single exposure.  The slit width was 1.5$''$ and
the slit length 13.6$''$, with a pixel scale of 0.8$''$ per pixel.  The
resolution averaged over the wavelength coverage of each grism is approximately
400.  Sky subtraction was achieved through observations of an offset blank sky
field after every target observation, with the exceptions of the planetary
nebulae, G35.20--1.74 and G43.89--0.78, where the objects were compact enough
that it was possible to nod along the slit.  Objects were aligned on the slit
using the imaging mode.  The co-ordinates given in Table 1 represent the
mid-point of the object as placed in the slit.

\begin{figure*}
\vspace{11cm}
\begin{minipage}{150mm}
\begin{centering}
\includegraphics{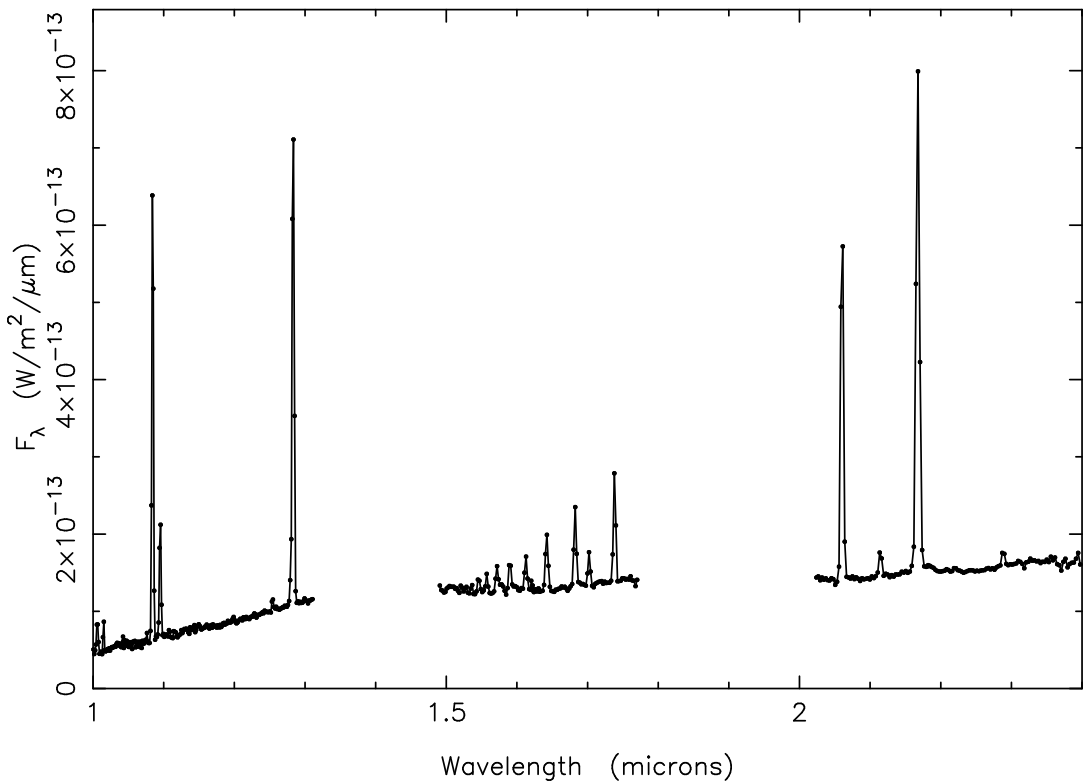}
\caption{The reduced 1.0 -- 2.8-$\mu$m spectrum of G298.23--0.33 containing
lines of the Paschen and Brackett series of hydrogen.  The three 
sections have been matched using a low-resolution grism spectrum as a 
template (see \S 2).
}
\label{g298specfig}
\end{centering}
\end{minipage}
\end{figure*}

The data were flat-fielded using a tungsten lamp before the appropriate on-off
pairs were combined.  Corrections for the curvature of the echelle orders were
then applied.  Finally spectra were extracted in a region covering the peak of
the emission and approximately 5 rows either side, or in the cases where sky
subtraction was derived from nodding along the slit, spectra were extracted at
the peak of the positive and negative images by summing over 3 pixel rows and
the two spectra were then subtracted.  Wavelength calibration was achieved by
means of an argon arc lamp.

The other observing mode used only the single-order H-grism.  This gave a
resolution of approximately 100 with the cs/36 secondary, and wavelength
coverage spanning 1.2--2.1$\mu$m.  These data were acquired to assist in
aligning the separate IJ and HK echelle spectra, since it was found that the
poor atmospheric transmission in the region 1.35--1.45$\mu$m, caused by water
and CO$_2$, made it difficult to join the {\em IJ} orders to the {\em HK}
orders accurately.  The low resolution spectra by comparison give a well
defined continuum shape covering wavelengths included by both echelles.  These
spectra were all acquired during the 1998 observations.  The same nod patterns
were used here as in the echelle observations.  Wavelength calibration was by
means of identifying the bright Paschen and Brackett lines in the planetary
nebulae.

The reduction of the straightened IRIS frames then followed a standard
procedure, using the {\sc figaro} data reduction package.  The same procedure
was applied to the standard-star spectra. Each object spectrum was then divided
by its corresponding standard spectrum, and corrected for the black-body shape
of the standard.

The numbers of lines detected depends on the brightness of the target, 
the extinction towards it and the atmospheric seeing.
Fluxes for the \hbox{H\,{\sc i}} emission lines were obtained by fitting
Gaussian functions.  Errors are taken from the fit parameters or assumed to be
5\%, reflecting the estimated accuracy of continuum matching between spectral
bands, whichever is the larger.

\section{Results}
\label{results}

Figure 1 shows the reduced JHK spectrum of G298.23--0.33 as an example
of the data obtained.  The separated continuum level has been matched using
the low-resolution grism spectrum as described above.

\begin{table}
\caption{H{\sc i} recombination-line line fluxes for the two planetary 
nebula targets, as a percentage relative to F(Br$\gamma$)}
\label{pntable}
\begin{tabular}{|c|c|c|c|r|}
Line       & $\lambda$(air)/$\mu$m & He2-96 & He2-117 & Case B\\
\hline
Pa$\delta$ & 1.005            & $81.5\pm6.0$  & $56.7\pm3.9$ & 201.1 \\
Pa$\gamma$ & 1.095            & $158\pm11$  & $121.2\pm8.5$ & 327.6 \\
Pa$\beta$  & 1.282            & $369\pm26$  & $329\pm24$ &   589.1 \\
Br16       & 1.556            & $5.8\pm0.4$  & $5.2\pm0.4$ & 8.0 \\
Br15       & 1.570            & $7.1\pm0.5$  & $6.9\pm0.5$ & 9.7 \\
Br14       & 1.588            & $9.3\pm0.7$ & $9.0\pm0.6$  & 11.9 \\
Br13       & 1.611            & $12.1\pm0.9$ & $10.9\pm0.7$  & 14.9 \\
Br11       & 1.681            & $19.6\pm1.4$  & $18.5\pm1.3$ & 24.7 \\
Br10       & 1.736            & $29.4\pm2.1$ & $27.1\pm1.9$  & 33.1 \\
Br$\gamma$ & 2.166            & $100\pm5$   & $100\pm5$  & 100.0 \\
\hline
\end{tabular}
\end{table}

\begin{figure*}
\vspace{17cm}
\includegraphics{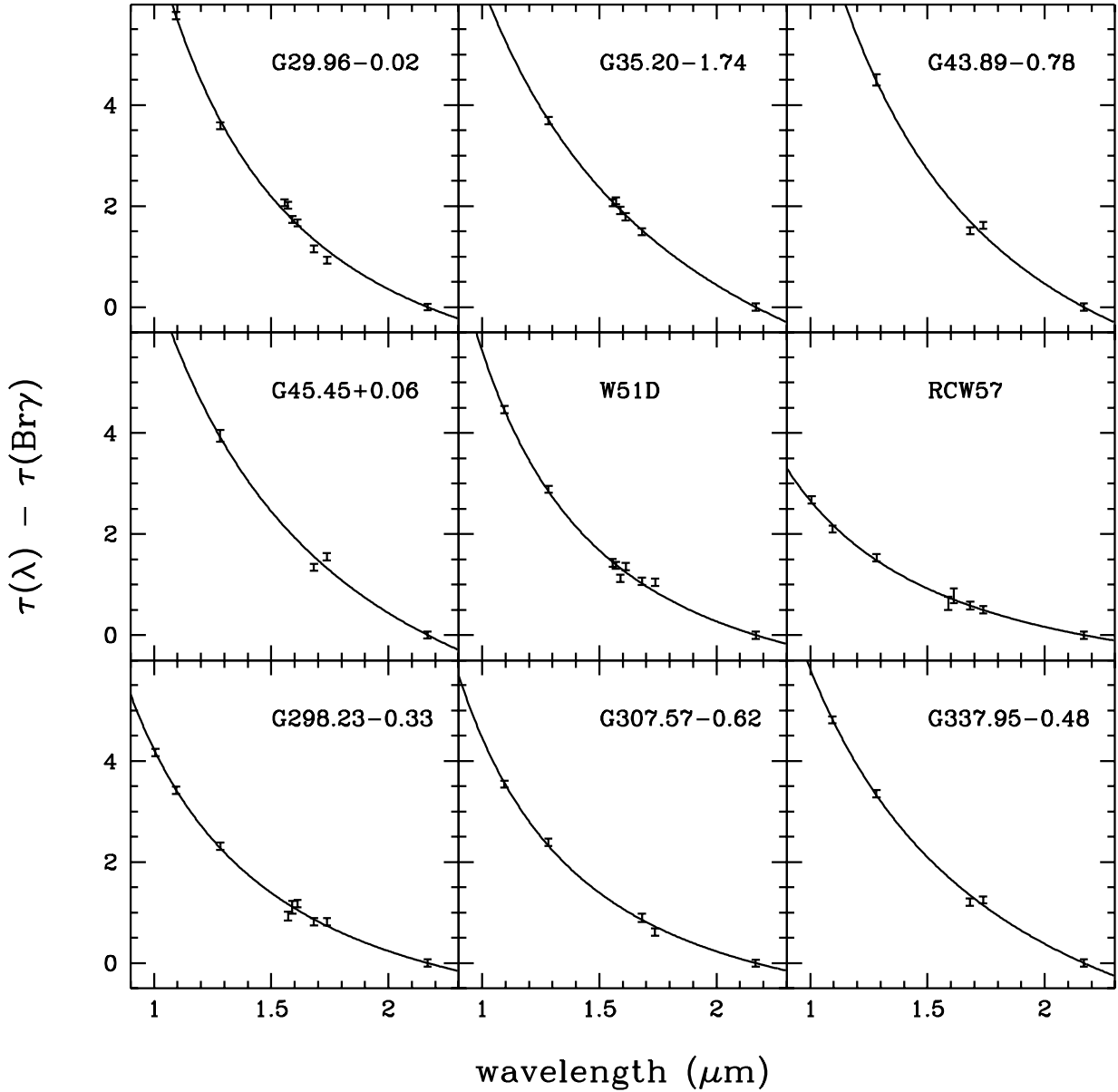}
\caption{Extinction-law fits to the H{\sc II}-region recombination-line data.
}
\label{hiifig}
\end{figure*}

\begin{figure}
\vspace{11.5cm}
\includegraphics{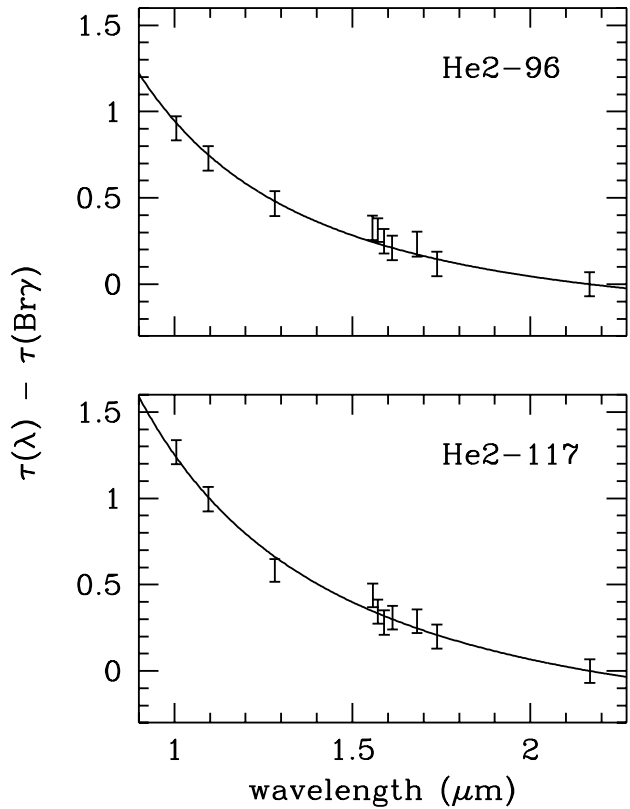}
\caption{Extinction-law fits to data for the PN He2-96 and He2-117.  The
numerical results are given in Table 2.
 }
\label{pnfig}
\end{figure}

\setlength{\tabcolsep}{4pt}
\begin{table*}
\caption{Measured hydrogen recombination-line fluxes for the \hbox{H\,{\sc ii}} regions,
as a percentage relative to F(Br$\gamma$).
}
\label{Hiilines}
\begin{tabular}{|c|c|c|c|c|c|c|c|c|c|}
Line       & G29.96--0.02 & G35.20--1.74 & G43.89--0.78 &
G45.45+0.06 & W51d & RCW\,57 & G298.23--0.33 & G307.57--0.62 & G337.95--0.48 \\
\hline
Pa$\delta$ & -- & -- & -- & -- &  -- &  $14.0\pm1.0$ & $3.1\pm0.2$ & -- & -- \\
Pa$\gamma$ & $1.03\pm0.07$ & -- & -- & -- &  $3.77\pm0.26$  &  $40.2\pm2.8$ & $1
0.7\pm0.7$  &  $9.5\pm0.7$ & $2.7\pm0.2$ \\
Pa$\beta$  & $16.3\pm1.1$   & $14.6\pm1.0$  & $14.2\pm1.6$  & $11.4\pm1.3$  & $3
2.9\pm2.3$ &  $127\pm9$  &  $58.1\pm4.1$  &  $53.9\pm3.8$  & $20.6\pm1.4$ \\
Br16     & $1.02\pm0.07$ & $1.02\pm0.07$  & -- & -- &  $1.92\pm0.13$ & -- & -- &
  -- &  -- \\
Br15     & $1.30\pm0.10$ & $1.18\pm0.08$  & -- & -- &  $2.43\pm0.17$ & -- &  $3.
8\pm0.3$ &  -- & -- \\
Br14     & $2.10\pm0.15$ & $1.75\pm0.12$  & -- & -- &  $3.87\pm0.27$ & $6.36\pm0
.83$ & $3.9\pm0.5$  &  -- & -- \\
Br13     & $2.81\pm0.19$ & $2.49\pm0.17$  & -- & -- &  $3.85\pm0.27$ &  $6.8\pm1
.0$ &  $4.6\pm0.3$ &  -- & -- \\
Br11     & $7.78\pm0.53$ & $5.57\pm0.39$  & $5.45\pm0.38$  & $6.46\pm0.45$ & $8.
53\pm0.60$  &  $13.7\pm1.1$ &  $10.9\pm0.8$ &  $10.1\pm0.8$  & $7.4\pm0.5$ \\
Br10     & $13.06\pm0.88$ & -- & $6.57\pm0.46$  & $7.00\pm0.50$  & $11.6\pm0.8$
 &  $20.0\pm1.4$ &  $14.5\pm1.0$ &  $17.9\pm1.3$ & $9.5\pm0.7$ \\
Br$\gamma$ & $100\pm5$   & $100\pm5$  & $100\pm5$  & $100\pm5$  &  $100\pm5$  &
$100\pm5$  &  $100\pm5$ & $100\pm5$  &$100\pm5$  \\
\hline
\end{tabular}
\end{table*}
\setlength{\tabcolsep}{6pt}
Hydrogen recombination-line flux ratios are given, relative to that of
Br$\gamma$, in Tables 2 and 3.  Table 2 contains results for the two
planetary nebulae, observed as low-extinction comparisons to the
H{\sc ii} regions.  These were acquired to test the accuracy of our 
observations since they have well defined optical extinction values.
Table 3 contains the main results for the H{\sc ii} regions.  The results 
are shown graphically in Figures 2 and 3 for the 
H{\sc ii} regions and planetary nebulae, respectively.
The Br12 line at 1.641\,$\mu$m has been excluded from all fits 
and calculations since it is blended in these observations with the 
[Fe{\sc ii}] 1.644\,$\mu$m line.

Observed line ratios are compared to predicted values from the
models of Hummer \& Storey (1995) with $n_e = 10^4$\,cm$^{-3}$ and
$T_e = 10^4$\,K.  Following Landini et al.\ (1984), if the wavelength
dependence of the dust opacity is assumed to follow a power law, so
that $\kappa(\lambda) \propto \lambda^{-\alpha}$, then the differential
optical depth between two lines, measured as

\[
\Delta \tau_{12} = -\ln \frac{\left[ I(\lambda_2)/I(\lambda_1) \right]_{\rm obs}
}{ \left[ I(\lambda_2)/I(\lambda_1) \right]_{\rm theory}},
\]

\noindent
is equivalent to

\[
\Delta \tau_{12} = \tau_{\lambda_1} \left[ (\lambda_2/\lambda_1)^{-\alpha} - 1
\right],
\]

\noindent
and the parameters $\tau_{\lambda_1}$ and $\alpha$ can be fitted to the data,
in this case, by minimising $\chi^2$ using the Marquardt method (Fortran
routine {\sc mrqmin} from Press et al.\ 1993).  We have
taken $\lambda_1 = \lambda$(Br${\gamma}$) = 2.166\,$\mu$m as the
reference wavelength in all cases.
The results of these fits are listed in Table 4 and plotted in Figures  
2 and 3.

\begin{table}
\caption{Extinction-law fit parameters}
\label{exttable}
\begin{tabular}{|c|c|c|c|c|}
Source &  $\alpha$  &  $\tau$(Br$\gamma$) & $\chi^2_{red}$ & $\tau_{10}$$^a$ \\
\hline
G29.96--0.02 & $1.90\pm0.05$ & $6.16\pm0.07$ & 4.71 & 1.2 \\
G35.20--1.74 & $1.11\pm0.12$ & $4.68\pm0.23$ & 0.48 & 4.6 \\
G43.89--0.78 & $1.63\pm0.14$ & $3.32\pm0.21$ & 6.5 & 2.8 \\
G45.45+0.06 & $1.33\pm0.16$ & $3.89\pm0.26$ & 10 & 2.7 \\
W51d & $1.95\pm0.07$ & $1.59\pm0.07$ & 2.57 & 1.7 \\
RCW\,57 & $1.39\pm0.14$ & $1.38\pm0.13$ & 0.57 & 1.7 \\
G298.23--0.33 & $1.59\pm0.08$ & $1.76\pm0.09$ & 1.86 & 1.7 \\
G307.57--0.62 & $1.79\pm0.14$ & $1.50\pm0.13$ & 1.3 & 2.3 \\
G337.95--0.48 & $1.23\pm0.09$ & $3.67\pm0.20$ & 1.9 & 2.5 \\
He2-96 & $1.93\pm0.32$ & $0.27\pm0.06$ & 0.49 & -- \\
He2-117 & $1.71\pm0.23$ & $0.46\pm0.07$ & 0.53 & -- \\
\hline
\end{tabular}
$^a$See \S4
\end{table}

\subsection{General Results}

Figure 4 shows the compiled results from fitting power-law extinction
curves to the data.  Also plotted is the result for G45.12+0.13 obtained by
Lumsden \& Puxley (1996).  Four objects have $\alpha$ values more than
2\,$\sigma$ below the mean interstellar value of 1.8 (e.g.\ Martin \&
Whittet 1990) and three of these are the three objects with largest
fitted values of $\tau$(Br$\gamma$).  None of the current targets has $\alpha$
significantly larger than 1.8.

The nine H{\sc ii} regions have 10-$\mu$m silicate optical
depths ($\tau_{10}$) estimated by Simpson \& Rubin (1990: hereafter SR)
or Faison et al.\ (1998), from fits to low-resolution spectra.
This gives us an independent check
of the apparent trend in Figure 4.  Figure 5(a) shows $\alpha$ plotted
against these, taking the lower of the $\tau_{10}$ values obtained by SR.
Figure 5(b) shows our fitted values of $\tau$(Br$\gamma$) plotted
against $\tau_{10}$.  

Five objects have $\tau_{10}$ values
and independent A$_V$ estimates available in the
literature.   These suggest that A$_V \sim 10 \tau_{10}$
and so $\tau_{10}$ should be comparable in size to the optical depth
at the wavelength of Br$\gamma$.


\begin{figure}
\vspace{8.5cm}
\includegraphics{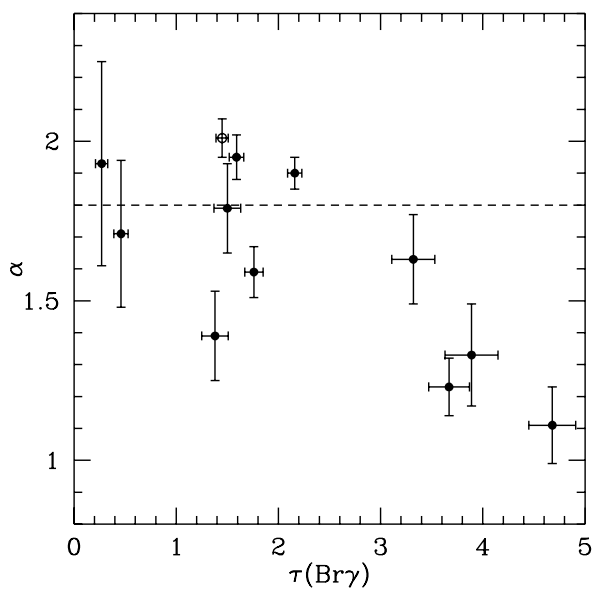}
\caption{Extinction-law fit parameters $\alpha$ (the power-law exponent)
against the continuum optical depth at Br$\gamma$.  The dashed line
represents $\alpha = 1.8$ the mean exponent for the general interstellar
medium (Martin \& Whittet 1990).  The open point represents the result
for G45.12+0.13 found by Lumsden \& Puxley (1996).
}
\label{alphatau}
\end{figure}
\subsection{Notes on individual objects}

\paragraph*{G29.96-0.02}
This object has a likely distance of 6\,kpc (Pratap, Megeath \& Bergin 1999).
Peeters et al.\ (2002) report no detection of 3.1-$\mu$m water-ice 
absorption in ISO data.
Estimates of extinction are A$_K$ = $2.14 \pm 0.08$
(Watson et al.\ 1997), A$_K$ = 1.6 and A$_H$ = 3.6 (Morisset et al.\ 2002).
Herter et al.\ (1981) estimate $\tau_{10} = 1.9 \pm 0.4$.
Faison et al.\ (1998) modelled $\tau(2.2) = 1.9$ and $\tau_{10} = 1.2$.
Our fit produces a $\tau$(Br$\gamma$) consistent with most of these,
and $\alpha$ about 2\,$\sigma$ steeper than the standard value of 1.8.

\paragraph*{G35.20--1.74}

At a distance of 3\,kpc (Chan, Henning \& Schreyer 1996), this source has
A$_V > 22$ mag (Woodward, Helfer \& Pipher 1985).  Our fitted $\tau$(Br$\gamma$)
is about twice the value extrapolated from this limit using a standard extinction 
law, and is therefore also consistent with the very flat curve we derive
(5--6\,$\sigma$ below the mean interstellar slope).
Faison et al.\ (1998) model $\tau(2.2) = 4.1$ and $\tau_{10} = 4.6$.
The 3.1-$\mu$m water-ice feature has been detected (paper in preparation).

\paragraph*{G43.89-0.78}

The estimated distance for this object is 9\,kpc and $\tau_{10} = 2.8$ (SR).  
Our fitted extinction curve is consistent with the standard law.

\paragraph*{G45.45+0.06}
$D = 4.2$ or 8.8\,kpc (Chan et al.\ 1996) and $\tau_{10} = 2.7$ (SR).
The mean A$_K \simeq 2.5$ across the H\,{\sc ii} region, with a peak
A$_K = 3.6$ (with corresponding A$_V$ values of 23 and 33: Feldt et 
al.\ 1998).  Our $\tau$(Br$\gamma$) is consistent
with the latter and $\alpha$ is about 3\,$\sigma$ flatter than the mean
extinction law.

\paragraph*{W51d}

The most luminous component of the giant \hbox{H\,{\sc ii}}
region W51, at a distance of 7.0\,kpc (Takahashi et al.\ 2000).
A$_V = 24\pm 3$ was determined by Goldader \& Wynn-Williams (1994).
Faison et al.\ (1998) modelled $\tau(2.2) = 2.6$ and $\tau_{10} = 1.7$.
Our value for $\alpha$ is 2\,$\sigma$ larger than the standard law.

\paragraph*{RCW\,57}

(G291.3--0.7) from the catalogue of Rodgers, Campbell \&
Whiteoak (1960). Its distance is given as 3.1\,kpc (SR).
Persson, Frogel \& Aaronson (1976) determined A$_V$ = 15 and SR
found a silicate optical depth of $\tau_{10} = 1.7$ (but see \S4).  
Our fit produces a value of  $\tau$(Br$\gamma$) consistent with these
and a rather flat $\alpha$ (3\,$\sigma$ below 1.8).
Pandey \& Ogura (1998) find $R = 4.4$ towards the
NGC\,3603 cluster, indicating that the extinction curve is also flatter
than the average at visible wavelengths.

\paragraph*{G298.23-0.33}

This object has estimated distance $\sim 11$\,kpc (Chan et al.\ 1996), A$_V 
\simeq 14$ (Persson
et al.\ 1976) and A$_K = 0.8$ (Mart\'\i n-Hern\'andez et al.\ 2002).
$\tau_{10} = 1.7$ (SR) but H$_2$O ice is not detected (Peeters et al.\ 2002).  
Our fitted
$\alpha$ is flatter than the interstellar mean by about 2\,$\sigma$.
$\tau$(Br$\gamma$) is about twice the above A$_K$ value, but is more
consistent with A$_V$ and $\tau_{10}$.

\paragraph*{G307.57-0.62}

The distance estimate is 2.7\,kpc (Chan et al.\ 1996) for this source
and $\tau_{10} = 2.3$ (SR).  Our fitted extinction 
law is consistent with the standard interstellar value.

\paragraph*{G337.95--0.48}

SR give $D = 2.9$\,kpc and $\tau_{10} = 2.5$ for this object.
Our spectra were taken during a period
of very poor seeing which has resulted in only a few of the Brackett
series lines being measured.   The estimate of $\alpha$ we
obtain is significantly (6\,$\sigma$) below the canonical
value.

\paragraph*{Planetary Nebulae He2-96, He2-117}

The derived extinction law for these objects is consistent with the
mean ISM law, although the error bars on our fits are large.  These
objects have measured Balmer decrements of $\sim 2.0$ (He2-96) and
2.7 (He2-117) (Tylenda et al.\ 1992).  Assuming the standard extinction
law, these values correspond to $\tau$(Br$\gamma$) = 0.33 and 0.44, 
respectively, consistent with our fitted values (Table 4). 

\begin{figure}
\vspace{11.5cm}
\includegraphics{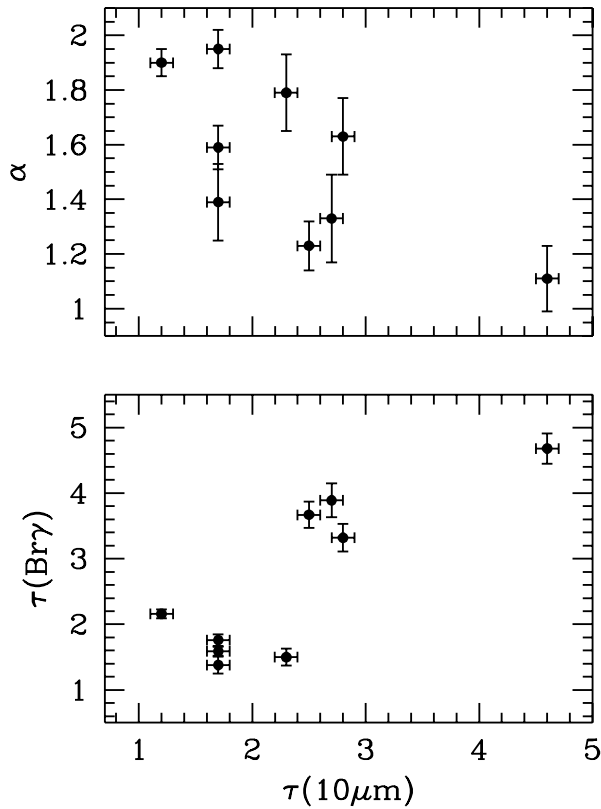}
\caption{(a) fitted values of $\alpha$ (the extinction-law exponent)
against the optical depth in the silicate feature $\tau_{10}$ listed
by Simpson \& Rubin (1990) and Faison et al.\ (1998); (b) fitted
$\tau$(Br$\gamma$) values against $\tau_{10}$.
Uncertainties in $\tau_{10}$ are taken to be 0.1 in each case.
}
\label{alphatau10}
\end{figure}

\begin{figure}
\vspace{11.5cm}
\includegraphics{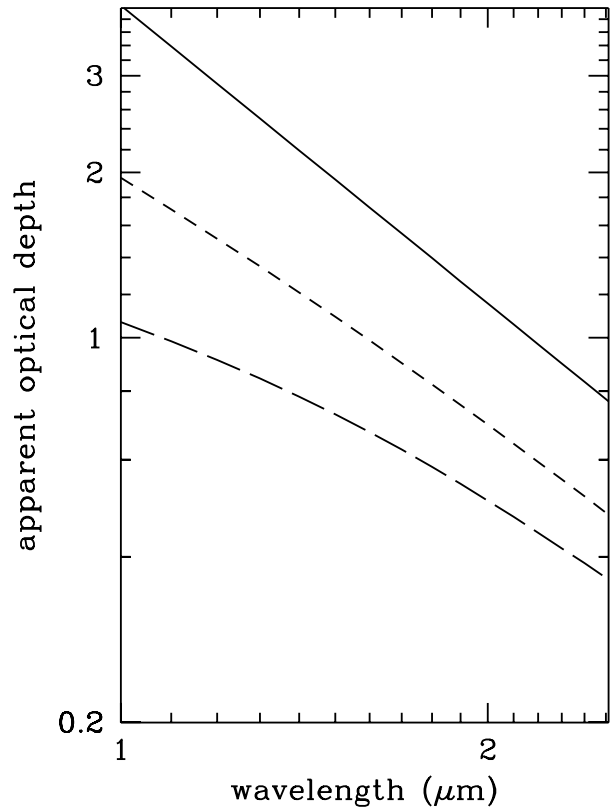}
\caption{A simple two-component model to demonstrate the apparent
flattening of the measured extinction law caused by unresolved optical depth
structure.  Here, there are two sources in the beam with optical depth
ratios of 1.0 (solid line), 3 (short dashed) and 10 (long dashed).
}
\label{structfig}
\end{figure}

\section{Discussion}
\label{discuss}

Figure 4 suggests an upper boundary to $\alpha$ and an increased
probability that objects with higher absolute extinction
($\tau$(Br$\gamma$) $\gs 2.5$ or A$_V \gs 25$) will have
1$\mu$m -- 2$\mu$m extinction curves that are flatter than the mean
interstellar law.  The data (including the PN and G45.12+0.13
points) have a Spearman rank correlation coefficient of
$r_s = -0.636$ which has an associated probability of arising at
random of $2.6 \times 10^{-2}$ for $N=12$.
This result lends support to the tentative report of flattened extinction 
for $A_V \gs 15$ by Kenyon, Lada \& Barsony (1998), from a statistical analysis 
of photometric data in $\rho$ Oph.  It also supports the finding of Racca 
et al.\ (2002) that flatter near-IR extinction laws are found where higher 
extinction regions are present.

Figure 5(a) shows an apparently similar distribution of
$\alpha$ with 10$-\mu$m silicate-absorption optical depth
obtained by fits to $IRAS$ LRS data (SR)
and to low-resolution ($R = 60$) spectroscopy (Faison et al.\
1998).  Figure 5(b) shows that our fitted optical depths vary
more or less as expected with $\tau_{10}$, although the significance
of the correlation is not high (the probability of it arising at 
random is 3.8 per cent).
A least-squares
fit gives a gradient of $1.0 \pm 0.3$ and intercept $0.3 \pm 0.7$.
The large scatter should not be too surprising, since there may be
large uncertainties in $\tau_{10}$ in these low-resolution spectra,
due to PAH emission features on either side of it.  There may
also be silicate emission components filling in the absorption feature
(Aitken 1996) and significant positional and beam-size discrepancies 
with our near-IR observations.

Figure 4 also shows that one target (RCW\,57) has a low fitted value
of $\alpha$ without having a particularly high extinction.  It should be
noted, however, that there are large amounts of
extended, structured emission (e.g.\ 2MASS K$_s$ images) in this
target and that it contains a cluster of objects, some of which have localised
high extinction 
(e.g.\ $\tau_{10} = 3.7$ or $A_V = 59$, Barbosa et al.\ 2003) close by
our observed position.  

There are two main potential causes of flatter observed extinction curves.
One is unresolved optical depth structure, the other is
a physical change in the dust grain properties, especially the size
distribution, in the material around the source.  These are discussed
below.

\subsection{Structure within the beam}

Figure 6 shows the results of a simple two-component model to demonstrate
the effects of extinction structure within the beam.  Three cases are
shown in which the wavelength-dependence of the extinction is $\tau(\lambda)
\propto \lambda^{-1.8}$ for both components, but where optical depths at
2.166\,$\mu$m are in the ratios 1.0, 3.0 and 10.0.  It can be
seen that the apparent extinction curve becomes flatter with increasing
contrast in the unresolved optical-depth structure, as well as departing
from a single power law.  Single power-law fits to the latter two
cases would produce exponents close to 1.5 and 1.2, respectively.
The cause of this artificial flattening is the penetration
of short-wavelength emission through the regions of low extinction.

As mentioned above, at least one H{\sc ii} region with a low fitted $\alpha$
value (RCW\,57) is associated with large amounts of 
extended structure.  However, 
there are two reasons to think that the effects of unresolved structure
may not be the principal
cause of the flatter extinction curves.   The first comes from the model,
which shows that the extinction contrast has to be large to produce curves
as flat as some of those observed.  Secondly, we might expect that
more distant sources have more unresolved structure and therefore
tend to have flatter extinction curves, but we find no correlation
between fitted $\alpha$ values and source distance.  We should, therefore
examine the effect of variable dust-grain properties.

\subsection{Grain properties}

\begin{figure}
\vspace{12.5cm}
\includegraphics{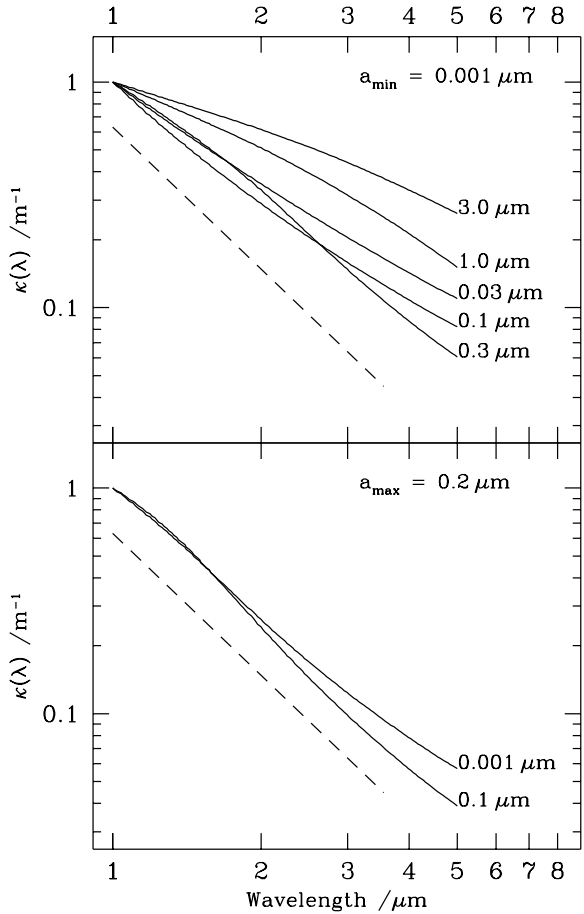}
\caption{Results of simple Mie-scattering model of the near-IR extinction
opacity due to spherical grains with a size distribution $n(a) \propto 
\lambda^{-3.5}$.
(a) shows the effect of increasing the maximum grain size; (b) shows the
that there is little effect in altering the minimum grain size.  The dashed
line represents a power law of $\lambda^{-1.8}$.  All models curves are
normalised to 1 at $\lambda = 1.0$\,$\mu$m.}
\label{K1fig}
\end{figure}

A simple model of extinction by spherical grains with a
size distribution similar to that deduced by Mathis, Rumpl
and Nordsieck (1977; MRN) can be used to place rough contraints on
the grain-size distribution in these regions.  We have calculated the opacity
of an equal-abundance mix of silicate and graphite grains using the Mie scattering
code {\sc bhmie} by Bohren \& Huffman (1983) and the optical constants used by
Weingartner \& Draine (2001).  We find that changes in the population
of smaller grains, as manifested in a variable minimum grain radius
$a_{\rm min}$ to the MRN-like distribution ($n(a) \propto a^{-3.5}$), does
not affect the slope of the near-IR extinction curve between 1$\mu$m
and 2$\mu$m (Figure 7).
The shape of the curve is much more sensitive to the maximum grain radius
$a_{\rm max}$ and becomes flatter both when $a_{\rm max}$ is small 
(because the absence of larger grains means that absorption becomes a 
more important component of the extinction)
and when $a_{\rm max}$ is large (when scattering by the larger grains
is no longer in the Rayleigh regime). Larger grains are the more likely 
case here, since there is other evidence of grain growth in star-formation
regions (see Introduction).

These simple models suggest that it is not easy to
obtain a near-IR extinction law that is significantly steeper than
the empirical power law ($\tau \propto \lambda^{-\alpha}$, $\alpha 
\simeq 1.8$).  A steeper slope would require
a large proportion of weakly absorbing grains in the mix.
Thick mantles of pure water ice (depth $\gs 0.1$\,$\mu$m) would have 
this effect, creating a steep extinction curve out to the edges
of the strong 3.1-$\mu$m feature (ignoring the weak features at
$\sim$1.5$\mu$m and $\sim$2$\mu$m).  Water ice exists
toward one source in our list (G35.20-1.74, section 3.2)
but, given that steep curves are not found, the mantles must be relatively
thin (significantly less than 0.1$\mu$m thick).  

The model results also imply that reproduction of the
standard extinction-law exponent, especially at wavelengths between
1\,$\mu$m and 2\,$\mu$m, requires values of $a_{\rm max}$ to be
between 0.1\,$\mu$m and 0.3\,$\mu$m (Fig.\ 7).  Physically, this limit
may arise from the balance between grain coagulation and destruction
processes.  Departures from it are more likely where there is high
line-of-sight extinction, but this is not the only variable (local density
and temperature should be involved).

There was little effect on the model predictions from introducing increases
in the silicate fraction, the addition of amorphous carbon or fluffy
composite grains, or from changes in the slope of the size distribution.

Our results are quite consistent with the predictions of the
model of Kr\"ugel \& Siebenmorgen (1994). These
authors also predict no significant change of slope of the extinction
curve between wavelengths of 1 and 2\um\ for grains with a
size distribution upper limit of 0.3\um,
applicable to dust in a dense molecular cloud.
(see their Figure 12). An observable flattening of the extinction law is
predicted by their models for grain
distributions with upper limits $>$ 0.3\um.
Other studies that fit grain models
to normal extinction also predict upper limits to
the grain radius in a narrow range close to 0.2\,$\mu$m (e.g.
Weingartner \& Draine 2001).

%


\section{Conclusions}

Determinations of the extinction law between 1$\mu$m and 2$\mu$m
have been made towards nine ultracompact H{\sc ii}
regions with A$_V \gs 15$, using H{\sc i} line ratios.  Four objects
have power-law fits $\tau \propto \lambda^{-\alpha}$ with fitted
values of $\alpha$ more than 2\,$\sigma$ lower than the average
interstellar value of 1.8.  None of the targets shows a significantly 
steeper extinction law.  There is an apparent trend for the sources
with highest extinction to have flatter extinction curves but low
values of $\alpha$ are found at all extinctions.  Although 
unresolved clumpy extinction structure can cause apparent flattening
of the extinction law, the observed shallow curves are likely to be
due to changes in the dust properties.  We will examine this
trend further in Paper II where we study the extinction law from 2--5$\mu$m 
in a sample of HII regions with average extinction higher than those 
presented here. A simple scattering model
shows that changes to the maximum grain radius in high column
density regions are the most likely cause and that the canonical
$\alpha = 1.8$ exponent suggests that the maximum grain size is
tightly constrained to be close to 0.2\,$\mu$m.  This upper limit
must be the result of balance between grain destruction and coagulation
processes.  Any significant change to the maximum grain radius due
to a change in this balance will cause a flattening of the near-IR
extinction curve.

\section*{Acknowledgments}

We thank the staff of the Anglo-Australian Observatory for their support
and the anonymous referee for helpful comments on the manuscript.

\end{document}